# Thermal Conductivity of Suspended Graphene with Defects

Hoda Malekpour[†], Pankaj Ramnani[‡], Srilok Srinivasan[§], Ganesh Balasubramanian[§], Denis L. Nika[†,∥], Ashok Mulchandani[‡], Roger Lake[¶] and Alexander A. Balandin[*,†]

[†]Phonon Optimized Engineered Materials (POEM) Center and Nano-Device Laboratory (NDL), Department of Electrical and Computer Engineering, University of California – Riverside, Riverside, California 92521 USA

[‡]Department of Chemical and Environmental Engineering, Bourns College of Engineering, University of California – Riverside, Riverside, California 92521 USA

[§]Department of Mechanical Engineering, Iowa State University, Ames, Iowa 50011, USA

[∥]E. Pokatilov Laboratory of Physics and Engineering of Nanomaterials, Department of Physics and Engineering, Moldova State University, Chisinau MD-2009, Republic of Moldova

[¶]Laboratory for Terascale and Terahertz Electronics (LATTE), Department of Electrical and Computer Engineering, University of California – Riverside, Riverside, California 92521 USA

## Abstract

We investigate the thermal conductivity of suspended graphene as a function of the density of defects, $N_D$, introduced in a controllable way. Graphene layers are synthesized using chemical vapor deposition, transferred onto a transmission electron microscopy grid, and suspended over ~7.5-μm size square holes. Defects are induced by irradiation of graphene with the low-energy electron beam (20 keV) and quantified by the Raman D-to-G peak intensity ratio. As the defect density changes from $2.0 \times 10^{10}$ cm$^{-2}$ to $1.8 \times 10^{11}$ cm$^{-2}$ the thermal conductivity decreases from ~$(1.8 \pm 0.2) \times 10^3$ W/mK to ~$(4.0 \pm 0.2) \times 10^2$ W/mK near room temperature. At higher defect densities, the thermal conductivity reveals an intriguing saturation behavior at a relatively high value of ~400 W/mK. The thermal conductivity dependence on defect density is analyzed using the Boltzmann transport equation and molecular dynamics simulations. The results are important for understanding phonon – point defect scattering in two-dimensional systems and for practical applications of graphene in thermal management.

**Content Image**

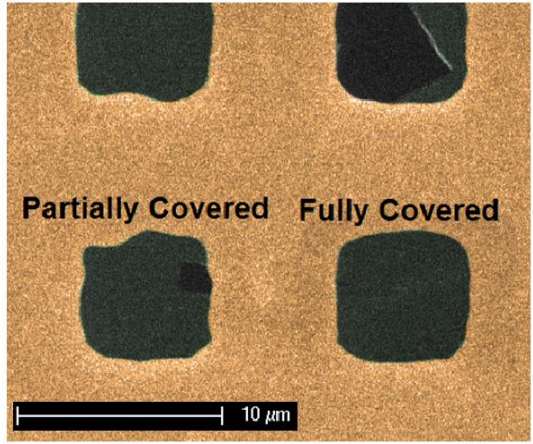
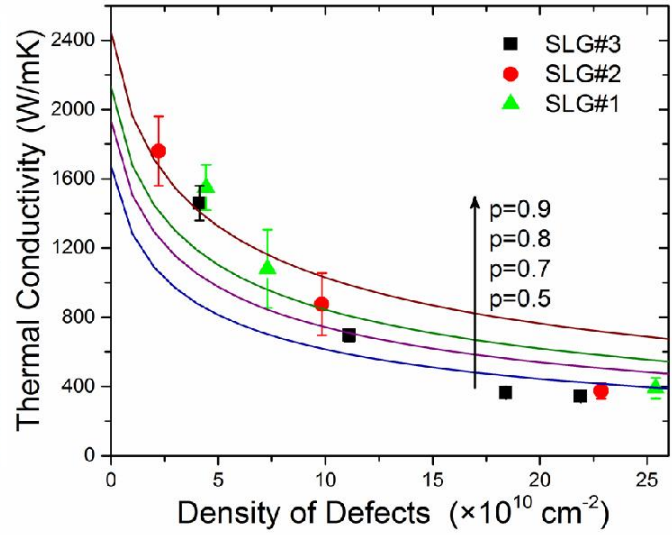

H. Malekpour, P. Ramnani, S. Srinivasan, G. Balasubramanian, D.L. Nika, A. Mulchandani, R. Lake and A.A. Balandin – 2016

Graphene [1] has exceptionally high intrinsic thermal conductivity, $K$ [2-3]. The measurements of thermal conductivity of large suspended graphene samples using the optothermal Raman technique revealed $K$ values exceeding those of bulk graphite, which is $K$=2000 W/mK at room temperature (RT) [3]. Independent measurements with the optothermal Raman technique [4-5] and the scanning thermal microscopy [6] confirmed the excellent heat conduction properties of graphene. Theoretical considerations suggest that graphene can have higher thermal conductivity than that of the graphite basal planes despite similar phonon dispersions and crystal lattice anharmonicities. The latter is attributed to an unusually long mean free path (MFP) of the long-wavelength phonons in two-dimensional (2-D) lattices [3, 7-8]. Recent calculations by different methods suggested that the graphene sample size should be in the 100 µm [9-10] or even 1 mm [11] range in order to fully recover the intrinsic thermal conductivity limited only by the lattice anharmonicity, i.e. without phonon scattering by defects, polycrystalline grains, and edges of the samples. The intrinsic $K$ values obtained in these works ranged from 4000 – 6000 W/mK near RT [9-11]. In other terms, the high intrinsic $K$ of graphene can be explained by the fact that the phonon Umklapp scattering is less efficient in restoring thermal equilibrium in 2-D systems than in bulk three-dimensional (3-D) systems [9, 12-13].

The thermal conductivity of graphene can be degraded by defects such as polymer residue from nanofabrication [14], edge roughness [8], polycrystalline grain boundaries [15], and disorder from contact with a substrate or a capping layer [16-18]. For this reason, the thermal conductivity of graphene synthesized by the chemical vapor deposition (CVD) is always lower than that of the mechanically exfoliated graphene from highly ordered pyrolytic graphite (HOPG) [2, 4, 19-21]. The loss of polycrystalline grain orientation in CVD graphene can lead to additional degradation of the thermal conductivity [22]. However, to date, there have been no quantitative experimental studies of the thermal conductivity dependence on the concentration of defects, $N_D$, in graphene. The knowledge of this dependence can shed light on the strength of the phonon – point defect scattering in 2-D materials. The change in the dimensionality results in different dependencies of the scattering rates on the phonon wavelengths in the processes of phonon relaxation by defects and grain boundaries [8, 23-24]. In bulk 3-D crystals, the phonon scattering rate on point defects, $1/\tau_P$, varies as $\sim 1/f^4$ (where $f$ is the phonon frequency) [24]. Owing to the changed phonon density of states (PDOS), the phonon scattering rate in 2-D graphene has a different frequency dependence,

H. Malekpour, P. Ramnani, S. Srinivasan, G. Balasubramanian, D.L. Nika, A. Mulchandani, R. Lake and A.A. Balandin – 2016

$1/\tau_P \sim 1/f^3$, which can, in principle, affect the phonon MFP and the thermal conductivity. In addition to the fundamental scientific interest, a quantitative study of the dependence of $K$ on $N_D$ is important for practical applications of graphene in thermal management. The graphene and few-layer graphene (FLG) heat spreaders [25-27] will likely be produced by CVD while FLG thermal fillers in thermal interface materials (TIMs) [28-30] will be synthesized via the liquid phase exfoliation (LPE) technique. Both methods typically provide graphene with a large density of defects.

In this letter, we report the results of an investigation of the thermal conductivity of suspended CVD grown single layer graphene (SLG) as a function of the density of defects, $N_D$. The unusual non-monotonic dependence of $K$ on $N_D$ is analyzed within the Boltzmann transport equation (BTE) approach and the molecular dynamics (MD) simulations. The samples for this study were grown by CVD on copper foils [31] and transferred onto gold transmission electron microscopy (TEM) grids with 7.5 μm × 7.5 μm square holes. Only the holes fully covered with graphene were chosen for the study to simplify the data extraction in the optothermal Raman technique [2, 32]. In this technique, a laser serves as source of heat and local temperature rise is determined from the shift in the Raman G peak position [2]. The technique and its modifications [4, 33] have been verified with other methods [34-35] and extended to a wide range of 2-D materials [36-37]. In the present measurements, the gold TEM grid (diameter – 3.05 mm, thickness ~ 25 μm) served as both the heat sink and the support for the suspended graphene, in a way similar to the experiments reported in Refs. [2, 4]. The high thermal conductivity of gold ($K$=350 W/mK) and strong attachment of graphene to gold grid ensured the accuracy of measurements. A scanning electron microscopy (SEM) image of the TEM grid is shown in Figure 1. The holes have black color while the gold parts appear yellow. One can notice that some holes are partially covered with graphene as seen from its greenish color. To ensure the accuracy and reproducibility of the results, the present study was conducted on three squares, which were completely covered with graphene.

[Figure 1]

The optothermal Raman technique is a non-contact steady-state technique, which directly measures the thermal conductivity [2-3]. The micro-Raman spectrometer acts both as a heater and thermometer. The measurement is done in two steps: the calibration procedure and the power-dependent Raman measurement. During the calibration, the Raman spectrum of graphene sample is recorded under low-power laser excitation in a wide temperature range [2]. In order to do this, the sample is placed inside a cold-hot cell (Linkam 600), where the temperature is controlled externally with steps of 10°C and accuracy of ~0.1°C. The samples are kept at least five minutes at each step to stabilize the temperature, and then the Raman G peak positions are recorded. The calibration Raman measurements are performed at low laser excitation power of ~0.5 mW to avoid any local heating caused by the laser. The procedure provides the position of the G peak as a function of the sample temperature. In the second step of the optothermal measurements, the excitation laser power is intentionally increased to cause local heating in the suspended graphene. The spectral position of the Raman G peak reveals the local temperature rise in response to the laser heating with the help of the calibration curve [2-3].

Figure 2 shows representative calibration (a) and power measurement (b) results. One can see from Figure 2 (a) that the dependence of the G peak spectral position on the sample temperature can be approximated as linear in the examined temperature interval. The extracted temperature coefficient $\chi_G$=-0.013 cm$^{-1}$/°C is in line with previous reports for graphene [38]. The G-peak shift with increasing laser power is presented in Figure 2 (b). One should note the excellent linear dependence of the G-peak shift on the laser power. The portion of light absorbed by suspended graphene, which causes the local heating, was measured directly by placing a power meter (Ophir) under the sample. To ensure accuracy, the absorbed power was measured for a graphene covered hole and on a reference empty hole. The difference in power readings corresponds to the power absorbed by graphene at a given laser wavelength $\lambda$. The measurement was repeated ten times at different laser power levels to determine the absorption coefficient of 5.68%±0.72% at the excitation laser wavelength of $\lambda$=488 nm. The light absorption coefficient at $\lambda$<500 nm, used in our experiments, is larger than the well-known long-wavelength limit, and it can increase further owing to surface contamination, defects, and bending [3, 39-41]. The slope of the $\omega_G(\Delta P)$ curve in Figure 2 (b) contains information about the value of thermal conductivity $K$, which can be extracted by solving the heat diffusion equation, knowing the sample geometry and temperature rise $\Delta T=\chi_G^{-1}\Delta\omega_G$ (where $\Delta\omega_G$ is the shift in the spectral position of G peak $\omega_G$). The large sample size ensures that the phonon transport is diffusive or partially diffusive. The details of the $K$ extraction procedure are provided in the Supplementary Information. The thermal conductivity of suspended CVD graphene before introduction of defects was found to be ~1800 W/mK near RT. This value is in agreement with the previous independent reports for suspended CVD graphene [4-5]. The grain boundaries, crystallographic misorientation, and defects reduce the thermal conductivity of CVD graphene as compared to that of graphene obtained by mechanical exfoliation from HOPG [2-3, 20-21].

[Figure 2]

The defects in the suspended graphene were introduced in a controllable way using low-energy electron beam irradiation [42-43]. The samples were exposed to 20 keV electron beams (SEM XL-30) with the beam current varying from ~ 3 nA to 10 nA. The irradiated area was kept constant at $6.6 \times 10^7$ nm$^2$ during the whole process. The irradiation dose was controlled by changing the beam current and irradiation time. The beam current was measured before each irradiation step using a Faraday cup. The details of the irradiation procedures are provided in the Supplementary Information. The Raman spectra of the suspended graphene samples were recorded after each irradiation step. Figure 3 shows the evolution of the D and D' peaks in the Raman spectrum of single layer graphene after each irradiation step. One can see from Figure 3 that the D to G peak intensity ratio, $I_D/I_G$, increases from 0.13 for as-grown CVD graphene all the way to 1.00, after four steps of the electron beam irradiation. The presence of the D peak in the spectrum before irradiation indicates a background defect concentration characteristic for CVD graphene and explains $K$ values somewhat below the bulk graphite limit [3, 44]. The evolution of the Raman spectrum under irradiation was used for quantifying the density of defects, $N_D$, following a conventional formula [45-46]:

$$N_D(cm^{-2}) = \frac{(1.8 \pm 0.5) \times 10^{22}}{\lambda^4}\left(\frac{I_D}{I_G}\right). \qquad (1)$$

[Figure 3]

It is known that Eq. (1) is valid for a relatively low defect-density regime with the inter-defect distance $L_D \geq 10$ nm. This criterion was met in the reported experiments. The defect density

H. Malekpour, P. Ramnani, S. Srinivasan, G. Balasubramanian, D.L. Nika, A. Mulchandani, R. Lake and A.A. Balandin – 2016

increases linearly with the Raman D to G peak intensity ratio. To show the correlation between the density of defects and the electron beam irradiation dose, we have plotted the Raman D to G peak intensity ratio, $I_D/I_G$, as a function of the total irradiation dose (see Figure 4). The linear dependence is clearly seen as expected for the low defect-density regime [42]. The optothermal Raman measurements were performed after each irradiation step. The temperature coefficient of the Raman G-peak, $\chi_G$, was not significantly affected by the defect density. In measuring the $\omega_G(\Delta P)$ dependence, we had to keep the power level small enough in order to avoid local healing of defects via heating.

[Figure 4]

In Figure 5 we present the extracted thermal conductivity, $K$, as a function of the defect density, $N_D$, by square, circle and triangle points corresponding to three suspended flakes of graphene. The details of the thermal data extraction have been reported by some of us elsewhere [7] and are briefly summarized in the Supplementary Information. For the small defect densities, $N_D < 1.2 \times 10^{11}$ cm$^{-2}$, the thermal conductivity decreased with increasing $N_D$. It can be approximated with the linear dependence $K = 1990 - 116 \times N_D$ [W/mK]. In the $N_D = 0$ limit, the thermal conductivity $K = 1990$ W/mK was still slightly smaller than that of the ideal basal plane of HOPG due to the background defects and grain boundaries present in CVD graphene before irradiation. At the defect density of $N_D \sim 1.5 \times 10^{11}$ cm$^{-2}$, one can see an intriguing change in the $K(N_D)$ slope followed by saturation. The thermal conductivity in the saturation regime was still rather high $K \sim 400$ W/mK. This is clearly above the amorphous carbon limit [3].

[Figure 5]

For theoretical interpretation of the measured behavior of the thermal conductivity we employed both a BTE analysis and MD simulations. The details of our BTE approach are provided in the Methods section. For analysis of our experimental data, we take into account three main mechanisms of phonon scattering: phonon – phonon Umklapp (U) scattering, phonon – rough edge scattering (also referred to as boundary (B) scattering), and phonon – point-defect (PD) scattering. Within the relaxation time approximation (RTA), the total relaxation rate is given as

H. Malekpour, P. Ramnani, S. Srinivasan, G. Balasubramanian, D.L. Nika, A. Mulchandani, R. Lake and A.A. Balandin – 2016

$$1/\tau_{tot}(s,q) = 1/\tau_U(s,q) + 1/\tau_B(s,q) + 1/\tau_{PD}(s,q), \tag{2}$$

where the index $s=LA, TA,$ or $ZA$ enumerates longitudinal acoustic (LA), transverse acoustic (TA), and out-of-plane acoustic (ZA) phonon polarization branches, and $q$ is the phonon wave number. The dependence of the thermal conductivity on the defect density, calculated from BTE within the RTA for different values of the specularity parameter $p$, is presented in Figure 5 by solid curves. The specularity parameter depends on the roughness of the edge and defines the probability of specular scattering of the phonons. For $p=1$, the scattering of phonons is purely specular, which means that the edge scattering does not introduce extra thermal resistance. For $p=0$, the scattering is fully diffuse, which corresponds to the strongest thermal resistance from the graphene edges [8-9].

The strength of the phonon scattering on defects is determined by the mass-difference parameter $\zeta=(\Delta M/M)^2$, where $M$ is the mass of carbon atom and $\Delta M=M-M_D$ is the difference between masses of a carbon atom and a defect. The value of $\zeta$ strongly depends on the nature of defects. In our BTE analysis, we used $\zeta$ as a fitting parameter to the experimental data. Within our model assumptions, the agreement with the experimental results is reached for $\zeta = 590$. The perturbation theory calculations [47] for pure vacancy defects in graphite estimate the value of the parameter to be $\zeta \sim 9$. This is substantially smaller than our fitting to the experimental data. The latter is explained by the fact that our model assumes only one type of phonon – defect scattering: mass-difference scattering on single vacancies. In reality, our samples contain a variety of defects, including those that were present before irradiation, e.g. grain boundaries, and those induced by irradiation, which are different from simple vacancies. Thus, large $\zeta$ imitates the effect of phonon scattering on all other types of defects. The important conclusion from the BTE modeling is that the saturation behavior can be reproduced via interplay of the three main phonon scattering mechanisms – Umklapp scattering due to lattice anharmonicity, point-defect mass-difference scattering, and rough edge scattering.

Let us now consider a possible nature of defects in our samples and their effect on the thermal conductivity as revealed from MD simulations. The details of our MD computational procedures are given in the Methods section. The electron energies of 20 keV used in the electron beam irradiation process are less than the knockout threshold energy of 80 keV [42; 48-50]. Such irradiation is only sufficient to overcome the energy barrier required for breaking of the carbon-carbon bond and initiating reaction with any residual impurities such as $H_2O$ and $O_2$ on the surface of graphene. This reaction results in functionalization of graphene with -OH and -C=O groups. Prior studies have shown that the –C=O configuration is energetically more favorable than –OH, and the transition of –OH and other functional groups into the energetically stable –C=O configuration can occur especially when they are annealed [51]. The energy barrier for the diffusion of -OH and epoxy groups is around 0.5-0.7 eV [52], which corresponds to a diffusion rate ~ $10^2$ $s^{-1}$ as calculated from transition-state theory, assuming a typical phonon frequency range in graphene. For this reason, the functional groups can be mobile at the temperature of the thermal experiments (~350 K). Upon continuous electron beam irradiation, two epoxy or hydroxyl group can come together and release an $O_2$ molecule [52]. When the coverage of functional groups is high, detectable amounts of $CO/CO_2$ can be released creating vacancies in the graphene lattice [53]. The presence of -OH and -C=O functional groups can be the reason for stronger phonon – defect scattering than that predicted by BTE models with vacancies only (and the resulting large $\zeta$ required for fitting to the experimental data). Our MD simulations show that a combination of single and double vacancy defects can also account for the experimentally observed saturation behavior of the thermal conductivity.

As one can see from Figure 6, the thermal conductivity decreases drastically for $N_D$ increasing from 2 $\times 10^{10}$ $cm^{-2}$ to $10 \times 10^{10}$ $cm^{-2}$ and subsequently saturates with the higher concentrations of defects. The $K$ saturation value is also substantially above the amorphous carbon limit – in line with the experiment. According to this model scenario, upon irradiation, -C=O and other functionalized defects are formed that strongly reduce the thermal conductivity. Continuous irradiation results in the creation of single and double vacancies. The increase in their concentration does not lead to pronounced $K$ reduction, which saturates to an approximately constant value for the $N_D$ range that was investigated. It can be explained in the following way. As more defects are introduced in graphene through irradiation the additional defect sites serve as scattering centers for phonons with wavelengths shorter than the distance between two vacancies. The delocalized long-wavelength phonons, that carry a significant fraction of heat, are less affected by extra defects that are closely spaced compared to those introduced at the previous irradiation step. At some irradiation dose, the increase in the phonon scattering rate of the delocalized modes due to extra defects is substantially smaller than that of the short-ranged localized modes. Hence, after a certain critical $N_D$ the thermal conductivity effectively saturates.

[Figure 6]

We further analyzed experimental Raman data to confirm the presence of vacancies in the irradiated graphene following the methodology developed in Ref. [54]. In this approach, the type of defects is determined from the ratio of intensities of D and D' peaks, I(D)/I(D'). It has been shown that that I(D)/I(D') in graphene attains its maximum ($\simeq 13$) for the defects associated with $sp^3$ hybridization, decreases for the vacancy-like defects ($\simeq 7$), and reaches a minimum for the boundary-like defects ($\simeq 3.5$) [54]. Following this method [54], the presence of vacancy type defects has been confirmed in our irradiated graphene sample (I(D)/I(D') $\simeq 7$). The details of this analysis are provided in Supplementary Information. It is known that the intensity of the D band depends not only on the concentration of defects [55], but also on the type of defects, and only defects that are capable of scattering electrons between the two valleys K and K' of the Brillouin zone can contribute to the D band [56-58]. For this reason, not all types of defects in graphene can be detected by Raman spectroscopy. However, our Raman data confirms the presence of vacancies supporting the theoretical assumptions.

In conclusion, we investigated the thermal conductivity of suspended CVD graphene as a function of the defect density. The defects were introduced by the low-energy electron beams and quantified by the Raman D-to-G peak intensity ratios. It was found that as the defect density changes from $2.0 \times 10^{10}$ cm$^{-2}$ to $1.8 \times 10^{11}$ cm$^{-2}$ the thermal conductivity reduces from $\sim (1.8 \pm 0.2) \times 10^3$ W/mK to $\sim (4.0 \pm 0.2) \times 10^2$ W/mK near RT. At higher defect density the thermal conductivity revealed an intriguing saturation behavior. The thermal conductivity reduction and saturation were explained theoretically within the Boltzmann transport equation and molecular dynamics approaches. The obtained results contribute to understanding the acoustic phonon – point defect scattering in 2-D materials. Our data indicating rather large values of thermal conductivity for graphene with defects adds validity to the proposed practical applications of graphene in thermal management.

H. Malekpour, P. Ramnani, S. Srinivasan, G. Balasubramanian, D.L. Nika, A. Mulchandani, R. Lake and A.A. Balandin – 2016

**METHODS**

**Graphene synthesis and transfer:** The single layer graphene samples were synthesized using ambient pressure chemical vapor deposition (AP-CVD) on a Cu foil. A polycrystalline Cu foil (99.8%, Alfa Aesar) was cleaned in acetic acid, acetone and IPA to remove any surface oxides. The cleaned Cu foil was loaded into the CVD chamber and the furnace temperature was ramped to 1030 °C while flowing Ar and $H_2$ and the foil was annealed for 2 h. For the growth of graphene, methane (90 ppm) along with Ar and $H_2$ was introduced into the chamber for 20 min. After the growth, the furnace was turned off and cooled to room temperature in Ar and $H_2$ atmosphere. Then, the SLG grains were transferred on to a gold TEM grid using a direct transfer method to avoid any contamination from the polymer support layer. A TEM grid (G2000, 7.5 μm square holes, TedPella) was placed directly on the Cu foil-graphene stack along with a drop of isopropyl alcohol (IPA). Upon heating, as IPA evaporates, the surface tension draws graphene and the metallic grid together into intimate contact. The Cu foil was then etched in ferric chloride, washed in DI water, and the resulting graphene on TEM grid was dried for use in subsequent Raman measurements.

**Electron beam irradiation:** The samples were irradiated under 20 keV electron beam using Philips XL-30 FEG field-emission system. The suspended graphene sample was exposed to continuous electron beam from electron gun with current varying from ~ 3 nA to ~ 10 nA controlled by the beam spot size. Before each irradiation step, the Faraday cup was used to read the beam current at the desired spot size. A constant magnification was maintained during all irradiation steps in order to keep the irradiated area constant ($6.6 \times 10^7$ nm$^2$). As a result, the dose density was controlled by the irradiation time. The irradiation process was done inside a vacuum chamber with the pressure below $10^{-4}$ Torr.

**Boltzmann transport equation approach:** In order to analyze the experimental data we used the BTE approach with the relaxation time approximation. In the framework of this BTE-RTA approach the thermal conductivity can be written as [8, 59]:

$$\kappa_G = \frac{1}{4\pi k_B T^2 h} \sum_{s=LA,TA,ZA} \int_0^{q_{\max}} \{[\hbar\omega_s(q)\frac{d\omega_s(q)}{dq}]^2 \tau_{tot}(s,q) \frac{\exp[\hbar\omega_s(q)/k_B T]}{[\exp[\hbar\omega_s(q)/k_B T]-1]^2} q\}dq, \qquad (3)$$

H. Malekpour, P. Ramnani, S. Srinivasan, G. Balasubramanian, D.L. Nika, A. Mulchandani, R. Lake and A.A. Balandin – 2016

where $h = 0.335$ nm is the graphene layer thickness, and the summation is performed over all acoustic phonon branches $s=LA$, $TA$ or $ZA$, $\omega_s$ is the phonon frequency of the $s$-th phonon branch, $q$ is the phonon wave number, $\tau_{tot}(s,q)$ is the total phonon relaxation time, $T$ is the absolute temperature, $\hbar$ and $k_B$ are Plank's and Boltzmann's constant, respectively. The scattering rates for the three main phonon relaxation processes, phonon-phonon Umklapp (U) scattering, phonon – rough-edge scattering (B) and phonon – point-defect (PD) scattering, are given by

$$1/\tau_B(s,q) = (\upsilon_s/L)((1-p)/(1+p)),$$
$$1/\tau_{PD}(s,q) = S_0 \Gamma q_s \omega_s^2 /(4\upsilon_s), \qquad (4)$$
$$\tau_{U,s} = \frac{1}{\gamma_s^2} \frac{M\upsilon_s^2}{k_B T} \frac{\omega_{s,\max}}{\omega^2}.$$

Here $\upsilon_s = d\omega_s/dq$ is the phonon group velocity, $p$ is the specularity parameter introduced above, $S$ is the surface per atom, $\omega_{s,\max}$ is the maximum cut-off frequency for a given branch, $\gamma_s$ is an average Gruneisen parameter of the branch $s$, $M$ is the mass of an unit cell, $\Gamma = \zeta(N_d/N_G)$ is the measure of the strength of the point defect scattering and $N_G = 3.8 \times 10^{15}$ cm$^{-2}$ is the concentration of carbon atoms.

**Molecular dynamics:** Simulations are performed on a pristine graphene sheet of size 319.5 nm × 54.1 nm containing 660,000 carbon (C) atoms. Defects (single and double vacancies) are introduced in the structure by randomly selecting and removing carbon atoms. The C-C interactions are described using the optimized Tersoff potential for thermal transport in graphene [60]. Periodic boundary conditions are employed in all directions. The simulations are carried out with the LAMMPS package [61]. The graphene structure is energy minimized and subsequently simulated under the isothermal-isobaric (NPT) ensemble using the Nose-Hoover thermostat at 300 K and barostat at 0 MPa for 4 ns, followed by equilibration in canonical (NVT) ensemble for 4 ns using the Nose-Hoover thermostat at 300 K. The coupling time for thermostats are 0.1 ps and that for barostat is 1 ps. The thermodynamic constraints are removed and the structure is simulated under the microcanonical (NVE) ensemble for 3 ns to ensure equilibration. Subsequently, thermal conductivity is computed using the reverse non-equilibrium MD technique [62, 63].

*Acknowledgements*

The work at UC Riverside was supported, in part, by the National Science Foundation (NSF) grant CMMI 1404967 on engineering defects in designer materials; NSF grant ECCS 1307671 on thermal properties of graphene, and by the Defense Advanced Research Project Agency (DARPA) and Semiconductor Research Corporation (SRC) via STARnet Center for Function Accelerated nanoMaterial Engineering (FAME). The work at Iowa State University was supported by the NSF grant CMMI-1404938 on engineering defects in designer materials. AAB acknowledges useful discussions with Professor A. V. Krasheninnikov on the nature of defects in graphene.

*Author Contributions*

A.A.B. initiated and coordinated the project, led experimental data analysis and wrote the manuscript; H.M. performed material characterization and Raman optothermal measurements; P.R. prepared suspended CVD graphene samples; A.M. supervised material synthesis and contributed to data analysis; D.L.N performed BTE simulation of heat conduction and contributed to data analysis; S.S. and G.B. performed the atomistic simulations and the corresponding thermal conductivity analysis; R.L. contributed to the computational data analysis. All authors contributed to manuscript preparation.

*Author Information*

The authors declare no competing financial interests. Correspondence and requests for materials should be addressed to (A.A.B.): balandin@ece.ucr.edu

***Supporting Information Available:*** The supporting information provides additional thermal conductivity measurements data. This material is available free of charge via the Internet at http://pubs.acs.org

**References**

[1] Geim, A. K.; Novoselov, K. S. *Nat. Mater.* **2007**, *6*, 183-191.

[2] Balandin, A. A.; Ghosh, S.; Bao, W.; Calizo, I.; Teweldebrhan, D.; Miao, F.; Lau, C. N. *Nano Lett.* **2008**, *8*, 902−907.

[3] Balandin, A. A. *Nat. Mater.* **2011**, *10*, 569−581.

[4] Cai, W.; Moore, A. L.; Zhu, Y.; Li, X.; Chen, S.; Shi, L.; Ruoff, R. S. *Nano Lett.* **2010**, *10*, 1645–1651.

[5] Chen, S.; Moore, A. L.; Cai, W.; Suk, J. W.; An, J.; Mishra, C.; Amos, C.; Magnuson, C. W.; Kang, J.; Shi, L.; Ruoff, R. S. *ACS Nano* **2010**, *5* (1), 321-328.

[6] Yoon, K.; Hwang, G.; Chung, J.; Kim, H. G.; Kwon, O.; Kihm, K. D.; Lee, J. S. *Carbon* **2014**, *6*, 77–83.

[7] Ghosh, S.; Bao, W.; Nika, D. L.; Subrina, S.; Pokatilov, E. P.; Lau, C. N.; Balandin, A. A. *Nat. Mater.* **2010**, *9*, 555–558.

[8] Nika, D. L.; Pokatilov, E. P.; Askerov, A. S.; Balandin, A. A. *Phys. Rev. B* **2009**, *79* (15), 155413.

[9] Nika, D. L.; Askerov, A. S.; Balandin, A. A. *Nano Lett.* **2012**, *12*, 3238-3244.

[10] Mei, S.; Maurer, L. N.; Aksamija, Z.; Knezevic, I. *J. Appl. Phys.* **2014**, *116*, 164307.

[11] Fugallo, G.; Cepellotti, A.; Paulatto, L.; Lazzeri, M.; Marzari, N.; Mauri, F. *Nano Lett.* **2014**, *14*, 6109–6114.

[12] Lepri, S.; Livi, R.; Politi, A. *Phys. Rep.* **2003**, *377*, 1−80.

[13] Nika, D. L.; Balandin, A. A. *J. Phys. Condens. Matter* **2012**, *24*, 233203.

[14] Pettes, M. T.; Jo, I.; Yao, Z.; Shi, L. *Nano Lett.* **2011**, *11*, 1195–1200.

[15] Serov, A. Y.; Ong, Z. Y.; Pop, E. *Appl. Phys. Lett.* **2013**, *102*, 033104.

[16] Saito, K.; Dhar, A. *Phys. Rev. Lett.* **2010**, *104* (4), 040601.

[17] Seol, J. H.; Jo, I.; Moore, A. L.; Lindsay, L.; Aitken Z. H.; Pettes, M. T.; Li, X.; Yao, Z.; Huang, R.; Broido D.; Mingo, N. *Science* **2010**, *328*, 213-216.

[18] Ong, Z. Y.; Pop, E. *Phys. Rev.* **2011**, *84* (7), 075471.

[19] Jauregui, L. A.; Yue, Y.; Sidorov, A. N.; Hu, J.; Yu, Q.; Lopez, G.; Jalilian, R.; Benjamin, D. K.;   Delk, D. A.; Wu, W. *ECS Trans.* **2010**, *28*, 73–83.

[20] Ghosh, S.; Calizo, I.; Teweldebrhan, D.; Pokatilov, E. P.; Nika, D. L.; Balandin, A. A.; Bao, W.; Miao, F.; Lau, C. N. *Appl. Phys. Lett.* **2008**, *92*, 151911.

[21] Li, H.; Ying, H.; Chen, X.; Nika, D.L.; Cocemasov, A.I.; Cai, W.; Balandin, A.A.; Chen, S.; *Nanoscale* **2014**, 6, 13402.

[22] Aksamija, Z.; Knezevic I. *Phys. Rev. B* **2014**, *90*, 035419.

[23] Klemens, P.G. J. *Wide Bandgap Mater.* **2000**, *7*, 332.

[24] Klemens, P. G.; Pedraza D. F. *Carbon* **1994**, *32*, 735.

[25] Yan, Z.; Liu, G.; Khan, J. M.; Balandin, A. A. *Nat. Commun.* **2012**, *3*, 827.

[26] Subrina, S.; Kotchetkov, D.; Balandin, A. A. *IEEE Electron Device Lett.* **2009**, 30, 1281.

[27] Gao, Z.; Zhang, Y.; Fu, Y.; Yuen, M. M. F.; Liu, J. *Carbon* **2013**, *61*, 342-348.

[28] Shahil, K. M. F.; Balandin, A. A. *Nano Lett.* **2012**, *12*, 861–867.

[29] Goyal, V.; Balandin, A. A. *Appl. Phys. Lett.* **2012**, *100*, 073113.

[30] Shahil, K. M. F.; Balandin, A. A. *Solid State Commun.* **2012**, *152*, 1331–1340.

[31] Zhang, Y.; Zhang, L.; Zhou, C. *Acc. Chem. Res.* **2013**, *46* (10), 2329-2339.

[32] Ghosh, S.; Nika, D.L.; Pokatilov, E.P.; Balandin, A.A. *New Journal of Phys.* **2009**, 11, 095012.

[33] Malekpour, H., Chang, K.H., Chen, J. C., Lu, C. Y., Nika, D. L., Novoselov, K. S. and Balandin, A. A. *Nano Lett.* **2014**, *14* (9), 5155-5161.

[34] Dorgan, V.E.; Behnam, A.; Conley, H.J.; Bolotin, K.I.; Pop, E. *Nano Lett.* **2013**, *13*, 4581–4586.

[35] Murali, R.; Yang, Y.; Brenner, K.; Beck, T.; Meindl, J. D. *Appl. Phys. Lett.* **2009**, *94*, 243114.

[36] Yan, R.; Simpson, J. R.; Bertolazzi, S.; Brivio, J.; Watson, M.; Wu, X.; Kis, A.; Luo, T.; Hight Walker, A. R; Xing, H. G. *ACS Nano* **2014**, *8* (1), 986-993.

[37] Yan, Z.; Jiang, C.; Pope, T. R.; Tsang, C. F.; Stickney, J. L.; Goli, P.; Renteria, J.; Salguero, T. T.; Balandin, A. A. *J. Appl. Phys.* **2013**, *114* (20), 204301.

[38] Calizo, I.; Balandin, A. A.; Bao, W.; Miao, F.; Lau, C. N. *Nano Lett.* **2007**, *7* (9), 2645-2649.

[39] Mak, K. F.; Sfeir, M. Y.; Wu, Y.; Lui, C. H.; Misewich, J. A.; Heinz, T. F. *Phys. Rev. Lett.* **2008**, *101* (19), 196405.

[40] Santoso, I.; Singh, R. S.; Gogoi, P. K.; Asmara, T. C.; Wei, D.; Chen, W.; Wee, A. T.; Pereira, V. M.; Rusydi, A. *Phys. Rev. B* **2014**, *89* (7), 075134.

[41] Ni, G. X.; Yang; H. Z.; Ji, W.; Baeck, S. J.; Toh, C. T.; Ahn, J. H.; Pereira, V. M.; Özyilmaz, B.; *Adv. Mater.* **2014**, *26* (7), 1081-1086.

[42] Teweldebrhan, D.; Balandin, A. A. *Appl. Phys. Lett*. **2009**, *94* (1), 013101.

[43] Liu, G.; Teweldebrhan, D.; Balandin, A. A. *IEEE Trans. Nanotechnol.* **2011**, *10* (4), 865-870.

[44] Ho, C. Y.; Powell, R. W.; Liley, P. E. *J. Phys. Chem. Ref. Data* **1972**, *1* (2), 279-421.

[45] Cançado, L. G.; Jorio, A.; Ferreira, E. M.; Stavale, F.; Achete, C. A.; Capaz, R. B.; Moutinho, M. V. O.; Lombardo, A.; Kulmala, T. S.; Ferrari, A. C. *Nano Lett*. **2011**, *11* (8), 3190-3196.

[46] Lucchese, M. M.; Stavale, F.; Ferreira, E. M.; Vilani, C.; Moutinho, M. V. O.; Capaz, R. B.; Achete, C. A.; Jorio, A. *Carbon* **2010**, *48* (5), 1592-1597.

[47] Ratsifaritana, C.A.; Klemens, P. G. *Int. J. Thermophys*. **1987**, *8* (6), 737-750.

[48] Kotakoski, J.; Krasheninnikov, A. V.; Kaiser, U.; Meyer, J. C. *Phys. Rev. Lett.* **2011**, *106* (10), 105505.

[49] Banhart, F.; Kotakoski, J.; Krasheninnikov, A. V. *ACS Nano* **2010**, *5* (1), 26-41.

[50] Feng, T.; Ruan, X.; Ye, Z.; Cao, B. *Phys. Rev. B* **2015**, *91* (22), 224301.

[51] Bagri, A.; Grantab, R.; Medhekar, N. V.; Shenoy, V. B. *J. Phys. Chem. C* **2010**, *114* (28), 12053-12061.

[52] Zhou, S.; Bongiorno, A. *Sci. Rep*. **2013**, *3*, 2484.

[53] Larciprete, R.; Fabris, S.; Sun, T.; Lacovig, P.; Baraldi, A.; Lizzit, S. *J. Am. Chem. Soc.* **2011**, *133* (43), 17315-17321.

[54] Eckmann, A.; Felten, A.; Mishchenko, A.; Britnell, L.; Krupke, R.; Novoselov, K. S.; Casiraghi, C. *Nano Lett.* **2012**, *12* (8), 3925-3930.

[55] Venezuela, P.; Lazzeri, M.; Mauri, F. *Phys. Rev. B* **2011**, *84* (3), 035433

[56] Cancado, L. G.; Pimenta, M. A.; Neves, B. R. A.; Dantas, M. S. S.; Jorio, A., *Phys. Rev. lett.* **2004**, *93* (24), 247401.

[57] Casiraghi, C.; Hartschuh, A.; Qian, H.; Piscanec, S.; Georgi, C.; Fasoli, A.; Novoselov, K. S.; Basko, D. M.; Ferrari, A. C. *Nano Lett*. **2009**, *9* (4), 1433-1441.

[58] Krauss, B.; Nemes-Incze, P.; Skakalova, V.; Biro, L. P.; Klitzing, K. V.; Smet, J. H. *Nano Lett.* **2010**, *10* (11), 4544-4548.

[59] Nika, D. L.; Ghosh, S.; Pokatilov, E. P.; Balandin, A. A. *Appl. Phys. Lett.* **2009**, *94* (20), 203103.

[60] Lindsay, L.; Broido, D. A. *Phys. Rev. B* **2010**, *81* (20), 205441.

[61] Plimpton, S. *J. Comput. Phys.* **1995**, *117* (1), 1-19.

[62] Müller-Plathe, F. *J. Chem. Phys.* **1997**, *106* (14), 6082-6085.

[63] Balasubramanian, G.; Puri, I. K.; Böhm, M. C.; Leroy, F. *Nanoscale* **2011**, *3* (9), 3714-3720.

H. Malekpour, P. Ramnani, S. Srinivasan, G. Balasubramanian, D.L. Nika, A. Mulchandani, R. Lake and A.A. Balandin – 2016

# Figures

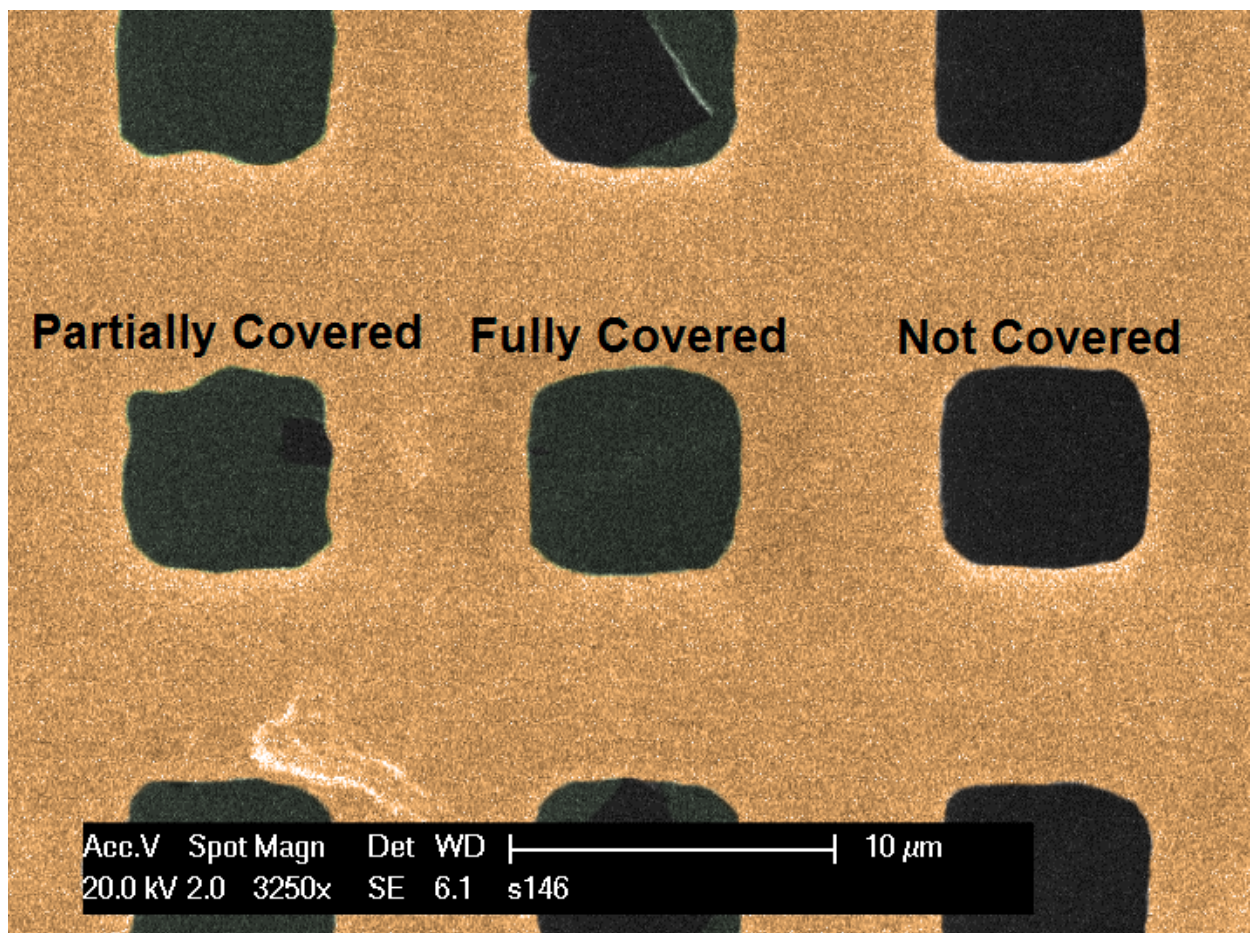

**Figure 1** Scanning electron microscopy image of graphene transferred on gold TEM grid showing 7.5-μm array of square holes. Some holes are fully or partially covered with the graphene flake. The grid is depicted in gold color, the holes are shown in black and the almost transparent greenish areas are suspended graphene flakes.

H. Malekpour, P. Ramnani, S. Srinivasan, G. Balasubramanian, D.L. Nika, A. Mulchandani, R. Lake and A.A. Balandin – 2016

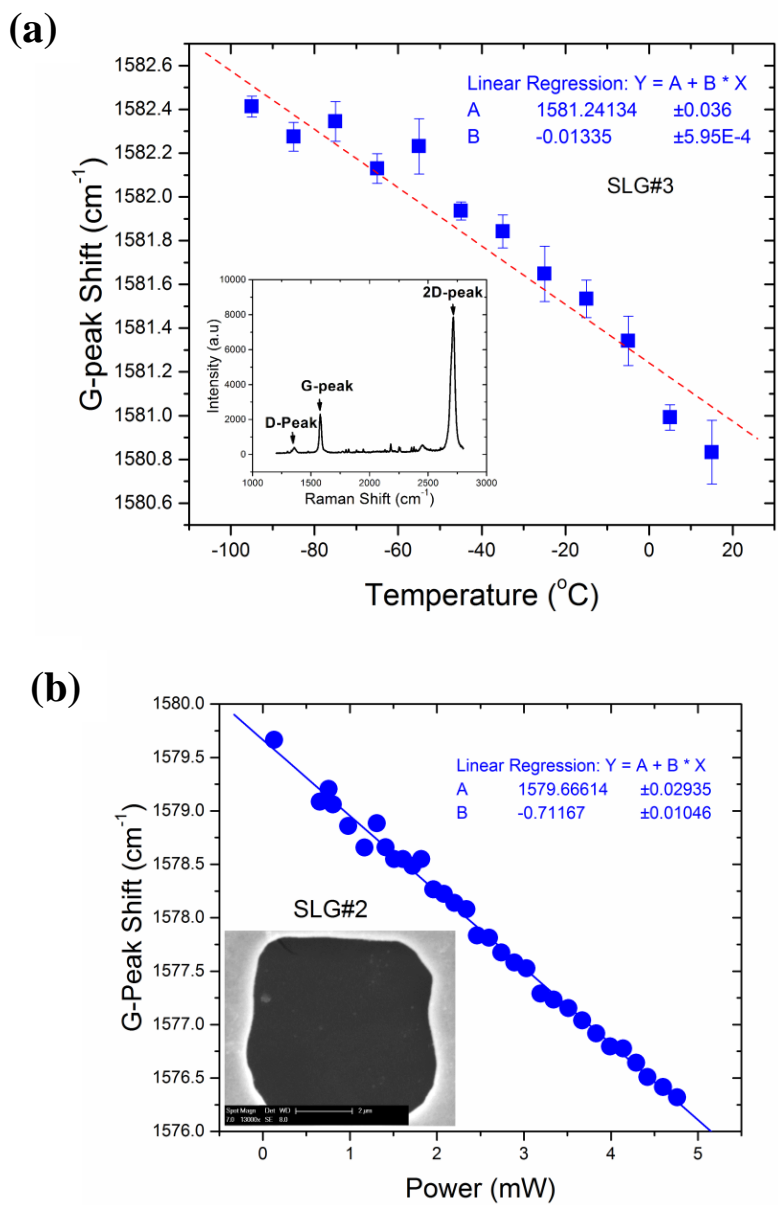

**Figure 2** Raman spectroscopy data for extraction of thermal conductivity of suspended CVD graphene flakes. (a) Calibration dependence of the Raman G peak position as a function of temperature. The measurement was conducted before graphene exposure to the electron beam. The inset shows a representative Raman spectrum of CVD graphene. (b) Raman G peak position dependence on the power on the excitation laser. The SEM image of this sample is depicted in the inset. The results demonstrate an excellent linear trend.

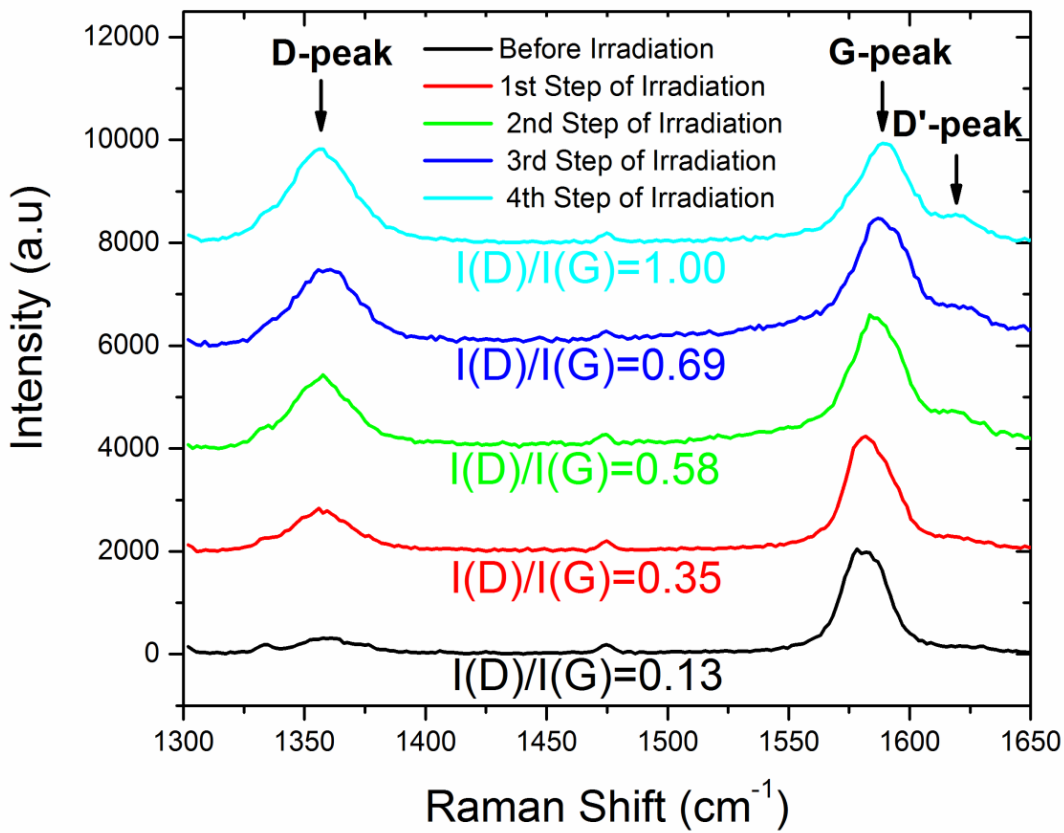

**Figure 3** Evolution of Raman spectrum under electron beam irradiation. As the sample is exposed to the electron beam, the Raman D peak intensity increases resulting in a D-to-G peak intensity ratio change from ~0.13 to ~1.00. The Raman G peak shifts to higher frequencies and the D' peak appears at ~1620 cm$^{-1}$. The Raman D to G peak intensity ratio is used to quantify the amount of induced defects.

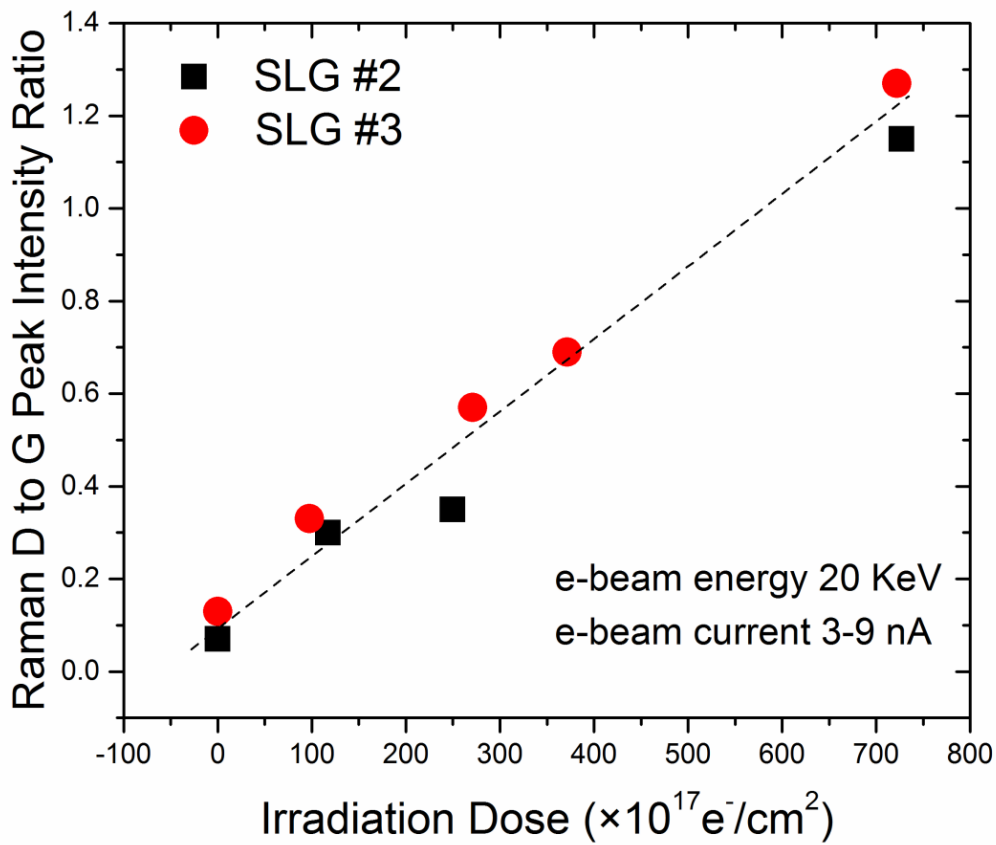

**Figure 4** Correlation of the Raman D-to-G peak intensity ratio with the electron beam irradiation dose. The low energy 20-keV electron beam was used to irradiate graphene. The beam current varied from ~3 to ~9 nA. The Raman D-to-G peak intensity ratio depends linearly on the irradiation dose.

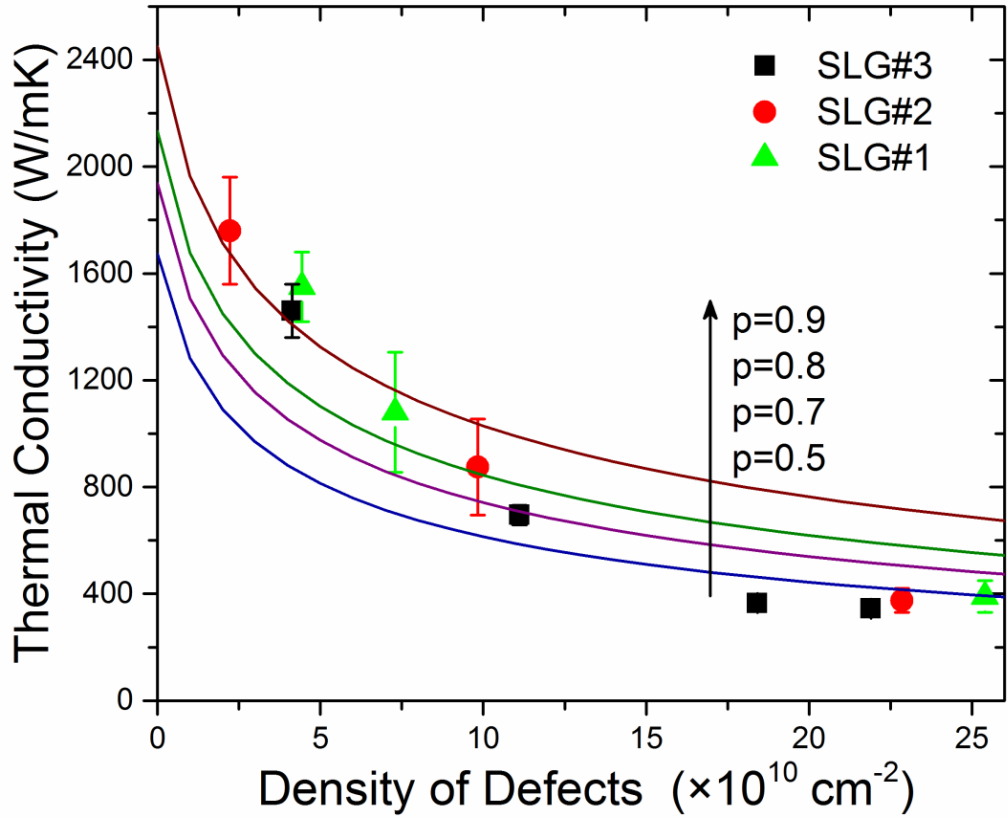

**Figure 5** Dependence of the thermal conductivity on the density of defects. The experimental data are shown by square, circle and triangle points. The solid curves are calculated using the BTE with different values of the specularity parameter *p*. Note that the interplay of three phonon relaxation mechanisms – Umklapp, point-defect, and rough edge scattering – gives a thermal conductivity dependence on the defect density close to the experimentally observed trend.

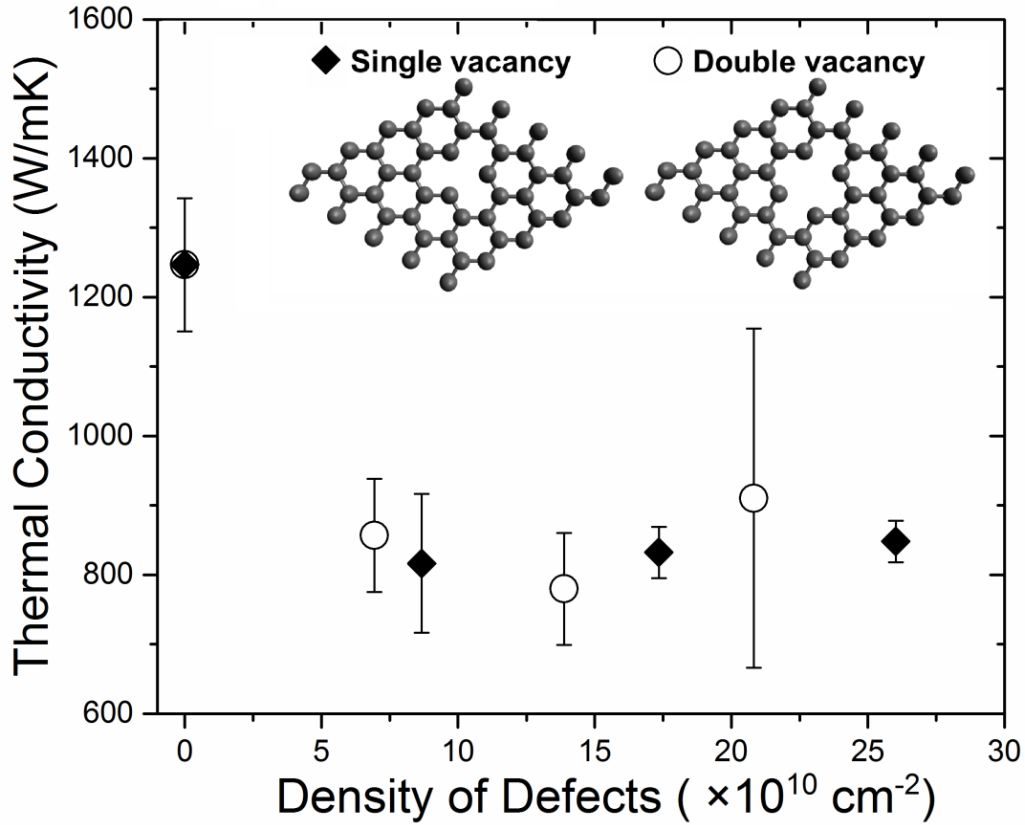

**Figure 6** Molecular dynamics simulation results for thermal conductivity of graphene with single and double vacancy defects. The simulated defect structures are depicted in the inset. The results show that the contributions of single and double vacancies are similar in reducing the thermal conductivity of graphene. The results are in line with the experimental trend.